# Temporal evolution of longitudinal bunch profile in a laser wakefield accelerator

M. Heigoldt[1,2], S. I. Bajlekov[3], A. Popp[1,2], K. Khrennikov[1,2], J. Wenz[1,2], S. W. Chou[1,2], B. Schmidt[4], S. M. Hooker[3] and S. Karsch[1,2]

[1]Ludwig-Maximilians-Universität München, Am Coulombwall 1, 85748 Garching, Germany

[2]MPI für Quantenoptik, Hans-Kopfermann-Str. 1, 85748 Garching, Germany

[3]Department of Physics, University of Oxford, Clarendon Laboratory, Parks Road, Oxford OX1 3PU, United Kingdom

[4]Deutsches Elektronen-Synchrotron DESY, Notkestrasse 85, 22607 Hamburg, Germany

**Due to their ultra-short duration and peak currents in the kA range[1,2], laser-wakefield accelerated electron bunches are promising drivers for ultrafast X-ray generation in compact free-electron-lasers (FELs), Thomson-scattering or betatron sources[3–5]. Here we present the first single-shot, high-resolution measurements of the longitudinal bunch profile obtained without prior assumptions about the bunch shape. Our method allows complex features, such as multi-bunch structures, to be detected. Varying the length of the gas target, and thus the acceleration length, enables an assessment of the bunch profile evolution during the acceleration process. We find a minimum bunch duration of 4.2 fs (full width at half maximum) with shot-to-shot fluctuation of 11% rms. Our results suggest that after depletion of the laser energy, a transition from a laser-**

**driven to a particle-driven wakefield occurs, associated with the injection of a secondary bunch. The resulting double-bunch structure might act as an elegant approach for driver-witness type experiments, i.e. allowing a non-dephasing-limited acceleration of the secondary bunch in a plasma-afterburner stage[6,7].**

Since the first demonstration of high-quality, quasi-monochromatic electron beams in 2004, laser wakefield acceleration (LWFA) has become a reliable scheme to accelerate electrons bunches to energies in the GeV range in plasma accelerator stages a few cm long[8–12]. The small scale of the acceleration structure, confining the bunch to a fraction of the plasma wavelength, implies bunch durations in the femtosecond range. Determining the detailed longitudinal profile of the generated bunches is important for understanding the accelerator dynamics, enabling accelerator control, and for determining their potential applications, such as driving compact FELs[13,14]. However, the limited temporal resolution of traditional methods, such as electro-optic sampling[15], prevents their application to measuring the ultra-short bunches produced by an LWFA. Although recent experiments confirmed the ultra-short nature of LWFA electron beams[1,2] these relied on the assumption of a Gaussian longitudinal profile when determining the electron bunch duration. As in earlier work[2], we determine the bunch profile from measurements of the spectrum of coherent transition radiation (CTR). However, our experiments advance prior work in several key aspects: (i) the bandwidth of the recorded spectrum covers a spectral range of more than 4 octaves at high resolution; (ii) the spectrum was recorded in a single-shot, preventing shot-to shot fluctuations in the electron bunch parameters distorting the measured spectrum; (iii) the

CTR spectrum was analyzed with a new algorithm (Bubblewrap)[16] which does not assume a form for the longitudinal bunch profile or extrapolation of the spectrum outside the measured range.

CTR is produced by the passage of a bunch of charged particles through the boundary between media with different dielectric indices. For a cylindrically symmetric bunch of $N_e$ electrons, the emitted energy W at frequency $\omega$ in observation direction $\theta$ is given by[17-19]:

$$\frac{d^2W}{d\omega d\Omega} = <\frac{d^2W_e}{d\omega d\Omega}> [N_e + N_e^2 |F(\omega, \theta)|^2],$$

where $<\frac{d^2W_e}{d\omega d\Omega}>$ is the weighted average of the single electron emission over the electron energy spectrum, and the formfactor $F(\omega, \theta) = \int \rho(\vec{x}) e^{-i\vec{k}\cdot\vec{x}} d^3x$ is the Fourier transform of the normalized three-dimensional bunch charge distribution $\rho(\vec{x})$ and $\vec{k}$ is the radiation wavevector in observation direction. When the bunch size is smaller than the radiated wavelength, transition radiation emitted from individual electrons is roughly in phase, leading to a coherent enhancement represented by the second term, while the first term describes the incoherent emission. For a relativistic beam with low divergence and no correlation between transverse and longitudinal distributions, the form factor may further be decomposed[19] into transverse and longitudinal components, i.e. $F = F_\perp F_\parallel$. If $F_\perp$ is known, the measured CTR spectrum directly yields the magnitude of the longitudinal form factor $F_\parallel$. Even then the longitudinal bunch profile $\rho_\parallel$ cannot be found just by inverse Fourier transformation of $F_\parallel$ since the phase information is not recorded.

An established reconstruction method often used in CTR experiments relies on the Kramers-Kronig relations to approximate the missing phase information[18], but this approach requires knowledge of the spectrum over the entire frequency domain. Since this is impossible, assumptions about the shape of the spectrum at low and high frequencies have to be made, and these can influence the deduced temporal profile[20]. In order to minimize the assumptions required, we have developed a new iterative algorithm (Bubblewrap) capable of reconstructing the longitudinal bunch profile. A detailed description of the algorithm is given in Ref[16]. Tests with synthetic data showed accurate reconstruction results, provided that the original data covered a sufficiently broad spectral range. In the experiments reported here the necessary spectral coverage was realized by combining two absolutely calibrated, commercial imaging spectrometers in the visible and near-infrared wavelength ranges with a third, custom-built spectrometer sensitive up to mid-infrared wavelengths, yielding a single-shot spectral coverage from 0.4-7.1 µm or 42-750 THz.

In this study, relativistic electrons with energies up to 650MeV were generated by focusing the driver laser into the entrance of a length-tunable, steady-state-flow gas cell filled with hydrogen. Forward CTR was generated by sending the bunches through a steel tape located 56 mm behind the gas cell entrance. The CTR was separated from the co-propagating electron beam by a reflective aluminium coated pellicle and analyzed by the spectrometers. The transverse electron beam size at the CTR radiator was 12-22 µm rms, depending on the length of the gas cell and correspondingly the distance between the gas cell exit and the fixed radiator foil (see Methods).

The experimental setup (see Fig 1) has the advantage that synchrotron radiation produced by the deflected electron beam does not reach the CTR diagnostics, but the disadvantage that the pellicle used for separating the CTR from the electron beam is itself a source of CTR. As described in the Supplementary Information, the spectral transmission function of the imaging system and the interference between CTR generated at the tape and pellicle were modelled by Fourier optics propagation, taking into account the near field distribution and relative phase delay. Although they were weak, interference effects were fully accounted for in our analysis. With the measured electron energy spectrum and inferred source size at both radiators, this procedure allows the CTR spectrum to be calculated in the detection plane for the case of full coherence ($|F_{||}(\omega)| = 1$). Dividing the measured CTR spectrum by this response function then yields $|F_{||}(\omega)|^2$, the square absolute value of the longitudinal formfactor. Our retrieval algorithm is then run on this formfactor to obtain the longitudinal bunch profile for each shot (see Fig 2).

In general, the charge contained within the reconstructed bunch profile is (80 +/-15)% of that measured independently by the electron spectrometer. While the deviation is still within the experimental error margin, it may be that the CTR measurements give a lower charge because a longer temporal feature produces CTR outside of our measurement range. Any such long temporal feature containing only a fraction of the measured charge would also exhibit a much lower peak current. Furthermore, we note that our analysis does not account for a possible chirp in the electron beam. Since the CTR emission increases with the electron energy our analysis always holds for the high-energy parts of the bunch, which are of most interest to applications. In case of a chirp, the additional delay would cause more destructive interference at higher frequencies compared to an un-chirped bunch, resulting in less bandwidth. Here our analysis would reconstruct a longer duration and thus give only an upper limit to the true pulse width.

The single-shot reconstruction technique was used in conjunction with the length tunability of the gas target to obtain the evolution of the bunch profile during the acceleration process. Data were taken at 1.0 mm intervals for target lengths in the range 3.0 – 14.0 mm; for each target length 30 consecutive shots were recorded.

For target lengths L <= 9.0 mm, the CTR spectra are smooth and consequently only a single electron bunch is observed, as shown in Fig 3. For a cell length optimized for maximum electron energy (L=9.0 mm), single electron bunches were generated with a maximum energy $E_{max}$ of 650 MeV, an average bunch duration of 4.2(±0.4) fs FWHM and a peak current of 5.4(±1.2) kA.

A different accelerator regime was observed for target lengths greater than 9 mm. For these conditions we observe a modulated spectrum, which immediately indicates that the temporal profile of the electrons contains more than one peak. The period $\Delta\omega$ of the spectral modulation immediately yields a peak separation of 15 µm which is close to one plasma wavelength ($\lambda_p \approx 17$ µm). Analysis with the Bubblewrap algorithm shows that these spectra are consistent with two distinct electron bunches separated by 15 µm, as shown in the plots for L > 9.0 mm in Fig.3. The mean charge contained in the 2nd electron bunch was observed to increase with cell length, reaching a maximum of 5pC at L =13 mm.

Complementary information about the underlying dynamics is provided by the evolution of the electron energies (Fig. 4). Maximum energies are reached at an acceleration length of L ≈ 9.0 mm. At approximately the same length (L = 8.0 mm) the onset of a 2$^{nd}$ energy peak is visible at the lower energy limit of our detection window, which we ascribe to the 2$^{nd}$ electron bunch. For L > 9.0 mm, the maximum electron energy remains approximately constant. This indicates that the acceleration process is not limited by de-phasing, which would cause reduced electron energy with increasing target length, but by laser pump depletion and/or diffraction. For our conditions, the pump depletion length[21] $L_{pd} \cong 8.7 * \frac{k_0^2}{k_p^3} \cong 10.4$ mm is in reasonable agreement with the experimental data.

With a transverse source size $\sigma_\perp \approx 0.95$ µm (determined with a similar setup to that employed in our previous work[22]), the charge density of the main bunch $n_b = N_e / (2\pi\sigma_\perp^2 \int \rho_\parallel(z)dz) \approx 2\times10^{19}$ cm$^{-3}$ is higher than the plasma electron density ($\frac{n_b}{n_e} \approx 5$). Further, the dimensions of the main bunch are smaller than the plasma wavelength ($k_p\sigma_\perp \approx 0.35$, $k_p\sigma_\parallel \approx 0.2$), such that the main bunch drives its own wakefield in the nonlinear, blowout regime of plasma wakefield acceleration (PWFA)[23,24]. Since the appearance of a 2$^{nd}$ electron bunch occurs approximately at the point of laser depletion, the transition from LWFA to PWFA mode is likely to be the trigger of the injection of the second bunch. Similar behaviour has been predicted by Pae et al.[25], who conducted particle-in-cell (PIC) simulations for similar electron beam parameters ($\frac{n_b}{n_e} = 4, \sigma_\perp \approx$ 2µm), and found that after the LWFA to PWFA transition, electrons were trapped and accelerated at the rear side of the beam-driven cavity. In those simulations, electrons appear to be injected from the remnants of the laser-driven wake upon its breakup after laser depletion. Additionally, during the transition between accelerator modes, the front of the plasma wave is retarded from its position during LWFA mode towards the electron bunch by approximately $\lambda_p/2$. The reduced plasma wave phase velocity during the mode transition may further assist an efficient trapping of non-relativistic plasma electrons[26,27].

In summary we have performed the first single-shot measurements of the longitudinal profile of electron bunches produced by a plasma accelerator and have performed the first measurements of the evolution of the bunch profile during acceleration; the results will find immediate application to studies of the use of plasma accelerators for driving compact FELs. The high temporal resolution and assumption-free bunch reconstruction provided by our technique allowed us to observe the complex dynamics of injection and acceleration at the interface between laser- and beam-driven particle acceleration. Our results are promising for so-called afterburner-acceleration[6,7], which would simplify the generation of high-quality electron beams from a hybrid accelerator by avoiding the need for an externally injected bunch from a prior stage.

**Methods:**

**Laser-plasma accelerator.** Experiments were performed using the ATLAS Ti:Sa laser system at the Max-Planck-Institut für Quantenoptik (MPQ) in Garching, Germany, which delivered 1.5(±0.1) J pulses of 28(±2) fs FWHM pulse duration on target. The beam was focused by an F/22 off-axis paraboloid into a length-tunable (3-14mm), steady-state-flow gas cell filled with hydrogen, resulting in a spot size of 18.7(±1.2)μm rms and a normalized vector potential of $a_0$ = 1.66(±0.13). Electron bunches were accelerated by means of self-injection in the weakly relativistic regime at a fixed plasma electron density of $n_e$ = 3.9x10$^{18}$cm$^{-3}$.

**Experimental Setup.** CTR was generated by the passage of the relativistic electron bunch through a pair of steel tapes (20 μm thick, 25 mm wide, 3 mm separation), located 56 mm behind the gas cell entrance. The tapes were advanced by a motor to provide clean material for each shot. The 1$^{st}$ tape blocked residual laser light and ensured that no light or thermal signal could reach the back of the 2$^{nd}$ tape. Forward CTR generated at the downstream side of the 2$^{nd}$ tape was separated from the co-propagating electron beam by a reflective Al-coated pellicle positioned 10 cm behind the tape and collimated by an f/3.75 off-axis paraboloid ($f_{eff}$= 19.05 cm). A 1 mm thick silicon wafer (HRFZ-Si, by Tydex) was used as a beam splitter to reflect ~50% of the CTR radiation through a BK7 window out of the vacuum chamber. Silicon is partially transparent at wavelengths below the bandgap (λ>1100 nm), acting as a low-pass filter for the transmitted CTR signal with a flat transmission of 50(±5)% in the relevant spectral range from 1.7 – 7.1 μm. The transmitted beam was directed to the mid-infrared spectrometer, directly connected to the main vacuum chamber. The part of the radiation reflected outside the experimental chamber was refocused by a lens (BK7, f=60 cm) and split by a 2$^{nd}$ silicon wafer of similar type onto the entrance slits of a near-infrared (Princeton Instruments OMA-C with 1024-element liquid nitrogen cooled InGaAs diode array, 1.1-1.8 μm), and visible (Oriel MS260, 420nm - 1100nm) spectrometer. The latter were absolutely calibrated, including the imaging optics, by usinga 1100K blackbody radiator (Lot-OrielLSB150), a tungsten halogen lamp (Ocean Optics HL-2000-CAL) and a He-Ne laser. We further confirmed that the measured signal is indeed CTR: When the plasma density was too low for wave-breaking to occur, no electron beam was generated and no CTR signal observed. When a band-pass filter (Si WBP, 8.0-14.0 µm, by Infratec) was placed in front of the mid-infrared spectrometer, the signal was limited to the accordant wavelength range, excluding the possibility of artefacts due to electromagnetic pulses generated by the accelerated electron beam.

The electron beam passing through the pellicle was dispersed by a dipole magnet and detected by a CCD-camera on an absolutely calibrated CAWO OG16 scintillator screen[28], allowing for determination of the charge and electron energy in the range between 200-1000 MeV. A second screen could be inserted in front of the dipole magnet to monitor the transverse profile and thus the divergence after 1.5 m of propagation. The transverse profile was found to be approximately Gaussian and the beam divergence independent of electron energy and gas cell length. Without the radiator foil we measured a beam divergence of 1.4(±0.3) mrad, yielding a transverse size of the electron bunch at the CTR radiator of 12-22 µm, dependent on the distance between the exit of the length-tunable gas cell and the fixed radiator foil.

**Mid-Infrared Spectrometer.** The custom built mid-infrared spectrometer is based on a design developed at DESY-FLASH[29]. Three consecutive gratings were installed, covering a range of 1.7-7.1 µm. Spectral intensity was recorded simultaneously by a total of 60 pyroelectric crystals (type LIM-107-X006 by InfraTec, LiTaO3 element). A detailed description of a similar instrument is given in Ref [29]. Initially, the spectral response of the instrument was calculated according to grating efficiencies (using the code MRCWA[30]) and wavelength acceptance $\Delta\lambda$ of each sensor. However, systematic variations in efficiency were found between individual pyro-elements, which typically caused the longitudinal bunch profile returned by the retrieval algorithm to contain small subsidiary bunches. Filtering these out and calculating the expected CTR spectrum generated by the dominant bunch allowed the smooth spectrum expected from a bunch to be calculated. Comparison of this smooth spectrum with the measured spectrum gave a correction factor for each pyro element, which was applied to all shots under all varying experimental conditions. Absolute efficiency was determined with a Nd:YAG laser.

**Retrieval Algorithm**. A detailed description of the algorithm is given in Ref[16]. The reconstruction grid uses N = $2^{12}$ points, spanning a frequency range of $\omega = 0 - 4.5 \times 10^{15}$ rad s$^{-1}$, which results in a data point spacing in the time domain of $\Delta t = 0.7$ fs.

**Acknowledgements:**

This work was supported by DFG through the MAP and TR-18 funding schemes, by EURATOM-IPP, the Max-Planck-Society, the Engineering and Physical Sciences Research Council (Grant No. EP/H011145/1) and the Leverhulme Trust (Grant No.F/08776/G).

**Author contribution**

M.H., S.B., A.P., B.S. and S.K. designed the experiment. M.H., S.B, A.P., K.K., J.W. and S.C. carried out the measurements. M.H., S.H. and S.K. wrote the main part of the paper. S.H. and S.K. provided overall guidance and supervised the project. All authors discussed the results, reviewed and commented on the manuscript.

**Author information**

Correspondence and request for materials should be addressed to S.K. (stefan.karsch@mpq.mpg.de).

**Figure legends**

Figure 1. Experimental set-up. The femtosecond laser pulse (red) is focused on the entrance of a length-tunable gas cell filled with hydrogen at a plasma electron density of $3.9 \times 10^{18}$ cm$^{-3}$. Residual laser light leaving the cell is blocked by two 20 μm thick steel tapes. Accelerated electrons (dark blue) leaving the plasma generate coherent transition radiation (light blue) when they traverse the tapes. Forward CTR produced at the backside of the 2$^{nd}$ tape is separated from the co-propagating electron bunch by an aluminium coated pellicle, collimated by an off-axis paraboloid, split by silicon wafers and directed into 3 optical spectrometers covering a spectral range from 0.4 – 7.1 μm. The electron beam energy and charge is analyzed by a pair of dipole magnets and a scintillating screen. Additionally a second scintillating screen (not shown) located in front of the magnets can be inserted to determine the beam divergence and transverse profile.

Figure 2. Measured and reconstructed longitudinal form factor $|F_{||}(\omega)|$ and reconstructed bunch profiles I(z) for 2 representative shots for gas cell lengths of 5 mm and 13 mm, respectively: (a,b) measured (black) and reconstructed (red) CTR spectrum, and reconstructed spectral phase (grey); Error bars show the rms error of the mid-infrared spectrometer (due to readout noise, uncertainty in spectral response of individual pyro-

electric detectors and transmission of the Silicon wafer). (c,d) Reconstructed longitudinal bunch profiles.

Figure 3: Bunch profile evolution: retrieved bunch profiles averaged over 30 consecutive shots for different length of the gas-cell. The grey band shows ±1σ, where σ is the shot-to-shot standard deviation for each 30-shot dataset.

Figure 4: Electron energy evolution: Measured electron spectra for gas cell lengths between 3 mm and 14 mm, recorded in 1 mm steps. For each length setting, 30 consecutive electron spectra are plotted as thin vertical lines. Red dots show the average cut-off energy (set to 10% of maximum spectral charge density) at each length; white dots show the average charge.

**Figures:**

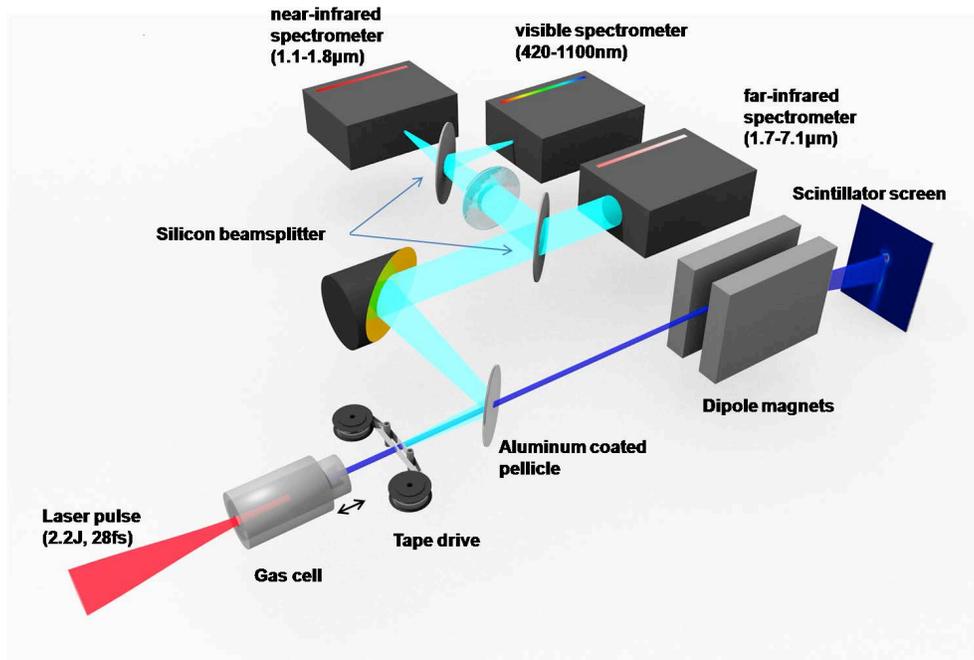

**Figure 1**

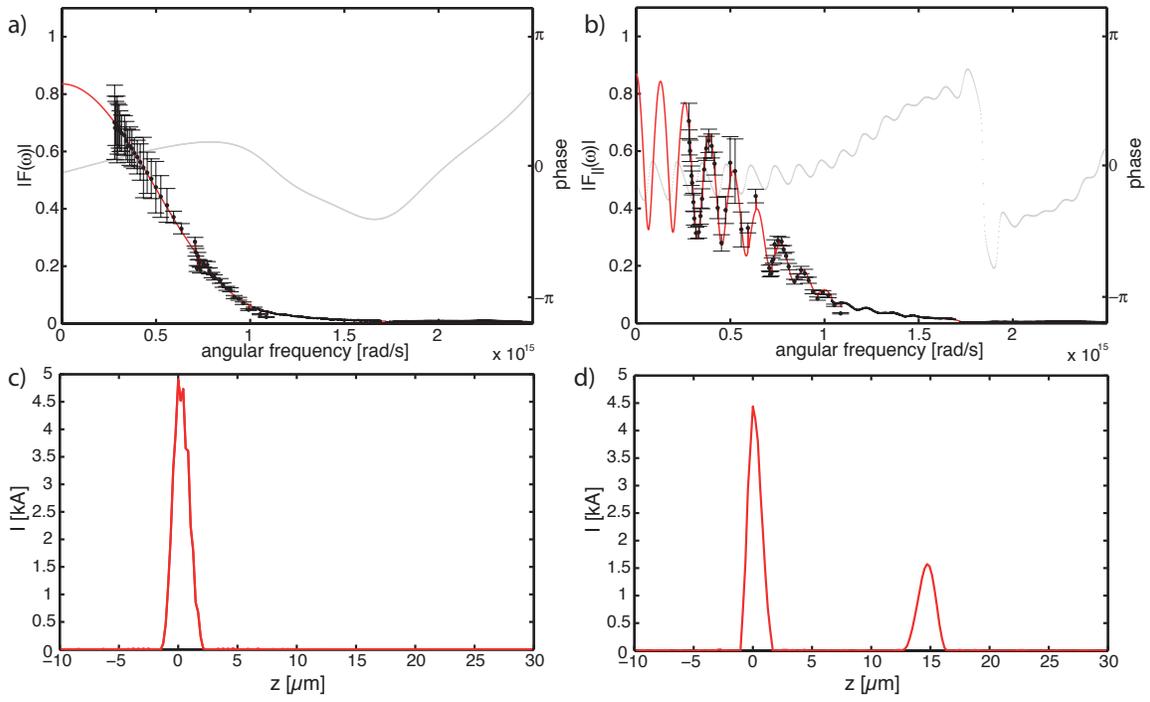

**Figure 2**

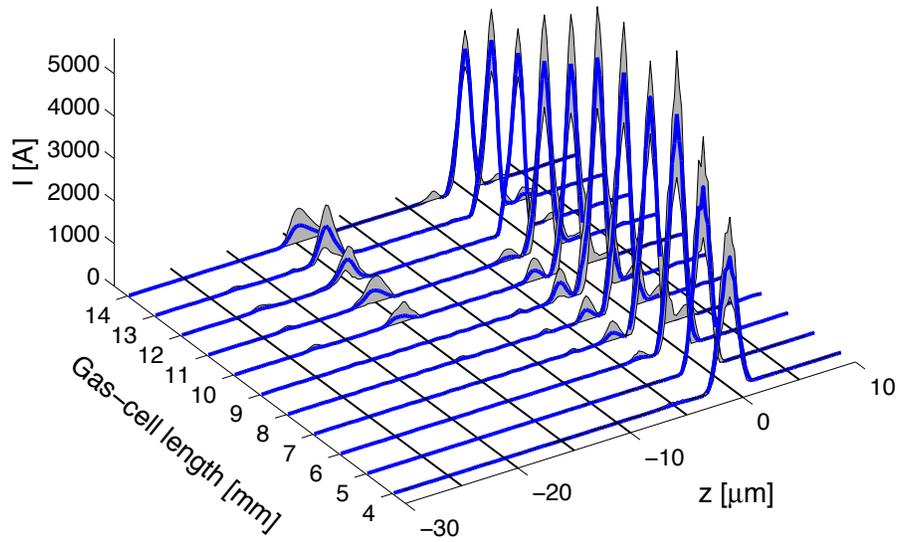

**Figure 3**

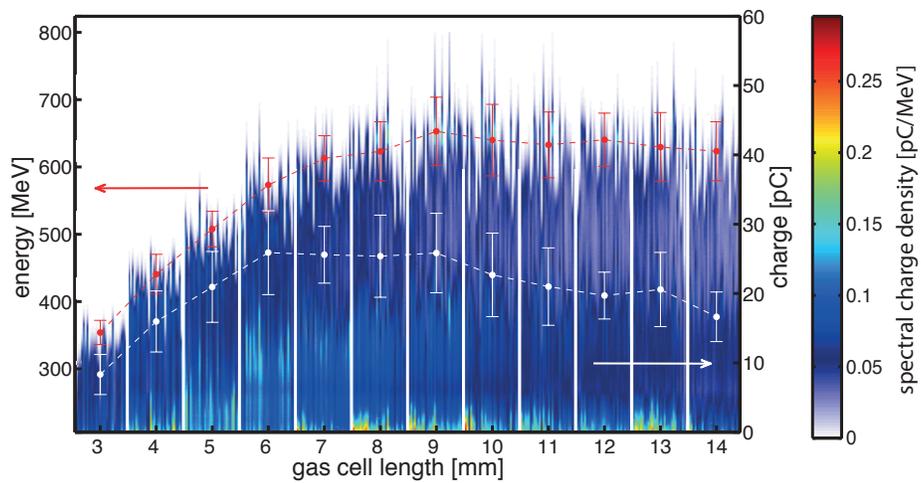

**Figure 4**

## Supplementary Information

### Spectral transmission function of the imaging system

Accurate measurement of the CTR spectrum relies on a thorough analysis of the CTR emission characteristics and imaging optics involved in the detection system. Furthermore, in the experimental setup used, CTR is not only produced at the tape but also at the reflective pellicle, which was used to separate the CTR radiation from the electron beam. Furthermore interference effects between both sources have to be considered, which could impact the spectrum being measured[1].

In the following, we describe the full Fourier optics transport calculation that was conducted to obtain the spectral transmission function of the detection system, as well as to quantify the contribution of both sources of CTR on the detected signal.

### Single electron emission

The angular spectral distribution of transition radiation in the far-field produced by a single relativistic electron is described by the Ginzburg-Frank formula,

$$\frac{d^2W_e}{d\omega d\Omega} = \frac{e^2}{4\pi^3\epsilon_0 c} \frac{\beta^2 \sin^2\theta}{(1-\beta^2\cos^2\theta)^2}$$

where $W_e$ is the emitted energy, $\omega$ is the angular frequency, $\Omega$ is the solid angle of collection, $\theta$ is the emission angle and $\beta = v/c$ is the normalized electron velocity. This formula is valid if the following relations between screen radius $r_{screen}$, observation distance D, the Lorentz factor of the electron $\gamma$ and the wavelength of transition radiation $\lambda$ are satisfied:

(1) $r_{screen} \geq \lambda\gamma$
(2) $D > L_f \approx \lambda^2\gamma$  (far-field condition)

Although, under our experimental conditions, relation (1) is always satisfied in the observed wavelength range, relation (2) is violated for long wavelength radiation and high electron energies, since the detection system is within the TR formation length $L_f$. This requires calculating the TR emission pattern for the near-field case and subsequent propagation through the imaging optics, preserving full phase information.

**Coherent transition radiation produced by an electron bunch**

The CTR emission produced by an electron bunch is given by the sum of the radiation field of each electron[2]. This summation can be removed by representing the electron beam by a six-dimensional distribution function $h(\mathbf{r}, \mathbf{p})$, where $\mathbf{r}$ and $\mathbf{p}$ are the position and momentum of an individual electron.

In the analytical model we use the following assumptions:

1. the electron kinetic energy is not correlated with its position in the bunch, such that $h(\mathbf{r}, \mathbf{p})$ is separable into a momentum distribution $g(\mathbf{p})$ and a spatial distribution $\rho(\mathbf{r})$, with $h(\mathbf{r}, \mathbf{p}) = g(\mathbf{p})\rho(\mathbf{r})$. Then, the angular spectral intensity of CTR is given by[3]:

$$\frac{d^2W_{CTR}}{d\omega d\Omega} = N_e^2 \int \frac{d^2W_e}{d\omega d\Omega}\, g(\mathbf{p})|F(\omega,\theta)|^2 d^3\mathbf{p} \qquad (3)$$

   where $N_e$ is the number of electrons contained in the bunch and $F$ is the spatial form factor defined by $F(\omega,\theta) = \int d^3\mathbf{r}\rho(\mathbf{r})\, e^{-i\mathbf{k}\cdot\mathbf{r}}$, and $\omega = c|\mathbf{k}|$.

2. there is no correlation between the radial and longitudinal spatial distributions, such that the spatial formfactor can be decomposed into its transverse and longitudinal component, $F_\perp$ and $F_\parallel$, with $F = F_\perp F_\parallel$ and

$$F_\perp(\omega,\theta) = \int d^2\mathbf{r}\, e^{-i\mathbf{k}_\perp \cdot \mathbf{r}_\perp}\rho(\mathbf{r}_\perp),$$

$$F_\parallel(\omega,\theta) = \int dz\, e^{-\frac{iz(\omega - \mathbf{k}_\perp \cdot \mathbf{v}_\perp)}{v_z}}\rho(z) \approx \int dz\, e^{-ikz}\rho(z) = F_\parallel(\omega).$$

   The last approximation is valid if the electron energies are relativistic and the beam divergence is small ($\Psi = \frac{v_\perp}{v_z} \ll 1$), which is the case in our experiment ($\Psi = 1.5\,\mathrm{mrad}$ FWHM).

Introducing the spectral transmission function of the detection system $T(\omega, \theta, p)$, and integrating over the solid angle of collection, equation (3) can be written as:

$$\frac{dW_{Detector,CTR}(\omega)}{d\omega} = N_e^2 |F_{||}(\omega)|^2 \cdot \left| \int d^3\mathbf{p}\, T(\omega, p) F_{\perp}(\omega) \frac{dW_e(\omega, p)}{d\omega}\, g(\mathbf{p}) \right|^2$$

Simple rearrangement yields:

$$|F_{||}(\omega)|^2 = \frac{dW_{Detector,CTR}(\omega)/d\omega}{N_e^3 \left| \int d^3\mathbf{p}\, T(\omega,p) F_{\perp}(\omega) \frac{dW_e(\omega,p)}{d\omega}\, g(\mathbf{p}) \right|^2} \qquad (4)$$

In the experiment, the electron beam momentum distribution $g(\mathbf{p})$ is measured and the transverse spatial size $\rho(\mathbf{r}_{\perp})$ is found to be well approximated by a Gaussian distribution. The only distribution not measured is $F_{||}$, which contains all information about the bunch temporal shape $\rho_{||}(z)$.

**Fourier propagation**

The denominator of equation (4) was obtained by first calculating the transmission function for each point in
the relevant experimental parameter space. This parameter space included TR frequency, electron energy, and transverse beam size. (The latter varied in a controlled manner for different lengths of the gas cell, due to the fact that the distance between the moveable gas cell exit and the tape changed, and hence the transverse beam size at the point of CTR emission.) The transmission function, along with the experimental parameters (electron energy spectrum, transverse size) and the measured CTR spectrum, allows us to obtain the absolute value of $F_{||}(\omega)$ for each shot, taking into account the contribution from both CTR sources (forward CTR produced at the tape and backward CTR produced at the pellicle).

Figure 1 shows the schematics of our optical path, along with the representation of the setup used in our model treatment.

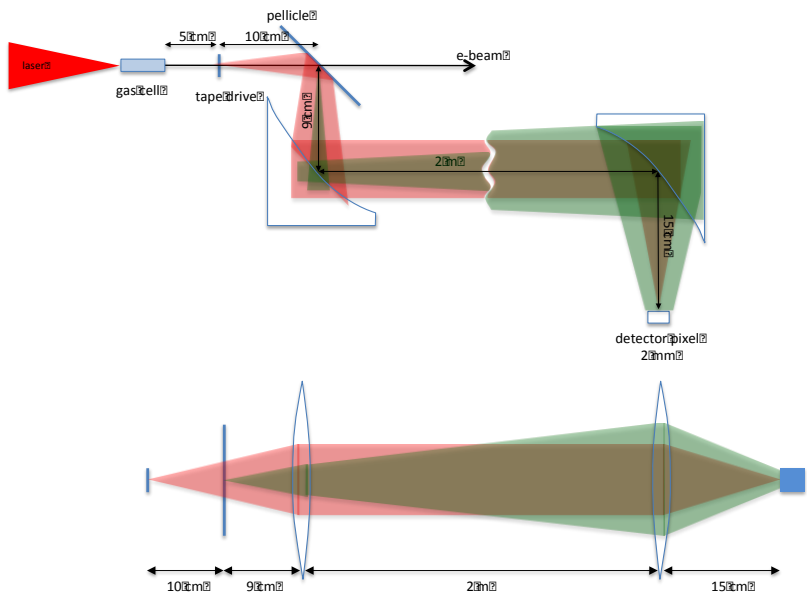

*Fig. 1: Schematic setup of the experiment (top) and modelling representation (bottom). In the case of the mid-infrared spectrometer, a grating (not drawn) in the 2 m propagation leg provides spectral separation to different detector pixels. The aluminium coated pellicle is roughly at half the focal distance from the first collimating optic, which leads to a large difference in wavefront curvature between the CTR beams from the tape and the pellicle. In the sketch, the divergences are strongly exaggerated.*

The calculation was performed for 64 evenly spaced frequency intervals (corresponding to a measurement range from 0.4 to 16 µm), 16 electron energies (in the detection range from 225-975MeV) and 11 different transverse sizes of the electron beam (corresponding to the discrete distances between the gas cell exit and the tape). The computation was done in 4 steps. In steps 1 to 3, the TR transmission was calculated for the relevant parameter space. In the last step, the experimental parameters where used to calculate the denominator of equation (4):

1. The spectral amplitude and phase of both CTR beams (produced at the tape and the pellicle)where computed at the entrance plane of the first lens for a mono-energetic electron sheet of unit charge, taking into account the transverse electron beam size at both radiators. The radiation field was calculated according to Ref[4], which is valid in the near-field. The mutual phase delay between both beams was taken into account, which is given by the velocity difference between electrons and TR radiation times the distance between tape and pellicle[1]. This step yields the (radially dependent) complex amplitude of both beams at the entrance plane of the first lens. The spectral amplitudes are plotted in Fig.2. Note that the radial polarization of CTR cancels the emission on axis.

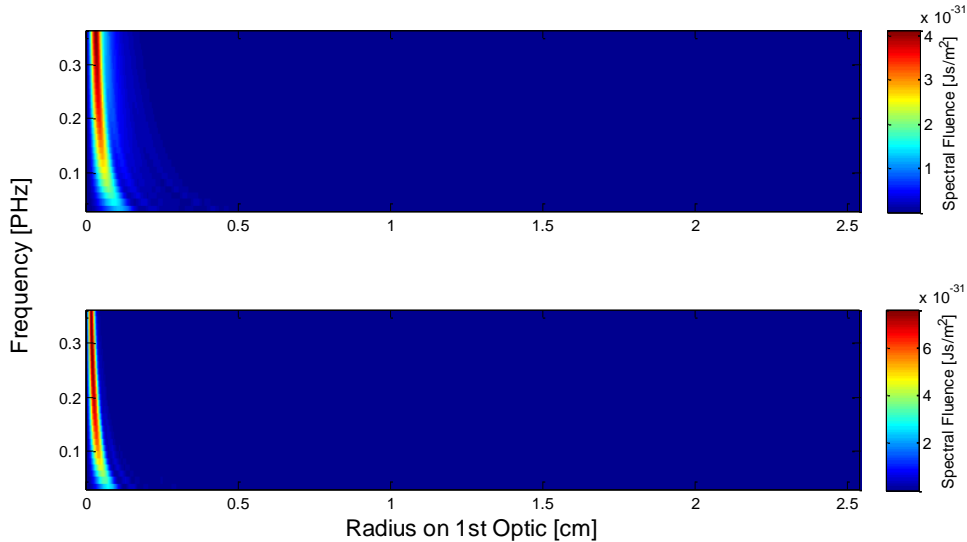

Figure 2: TR spectral/radial distributions in the entrance plane of the first collimating optic. Top: beam from tape drive. Bottom: beam from pellicle. Electron beam parameters: E = 525MeV, FWHM source radius of 0.95µm and a divergence of 1.5mrad, crossing the tape after 50mm of propagation.

2. After calculating the wavefront modification introduced by the first parabola, both beams are propagated over 2m using electromagnetic wave propagation by Fourier transformation[4,5]. This yields the complex amplitudes at the entrance plane of the second parabola. Figure 3 displays the spectral/radial distribution at the entrance plane of the second (focusing) optic.

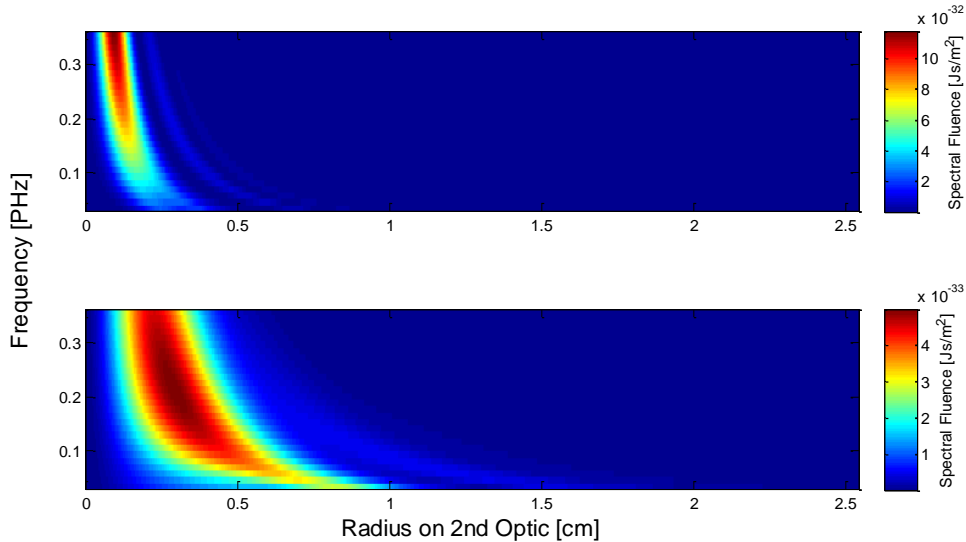

*Figure 3: TR spectral/radial distributions after 2m propagation in the entrance plane of the focusing optic. Top: beam from tape drive. Bottom: beam from pellicle. Despite the fact that the first optic acts as a collimator for the beam from the tape (top), at wavelengths of a few microns diffraction leads to a significant increase of the beam size during the propagation over 2 m. The beam from the pellicle spreads even further, because it is not collimated by the first lens.*

3. A further beam propagation of the complex amplitude after the second lens yields the electric field distribution for both beams in the focal plane (f=15 cm) of the focusing optic. Plotted in Fig. 4 are radial lineouts of the focus distributions. While the beam from the tape is well focused onto the detector, the beam from the pellicle forms a much larger spot due to its lack of collimation. The region of radii where interference can occur is limited to the very outer parts of the beam from the tape and inner parts of the beam from the pellicle. Due to the large difference in size, interference between both beams has only a small effect on the measurement (see Fig. 6).

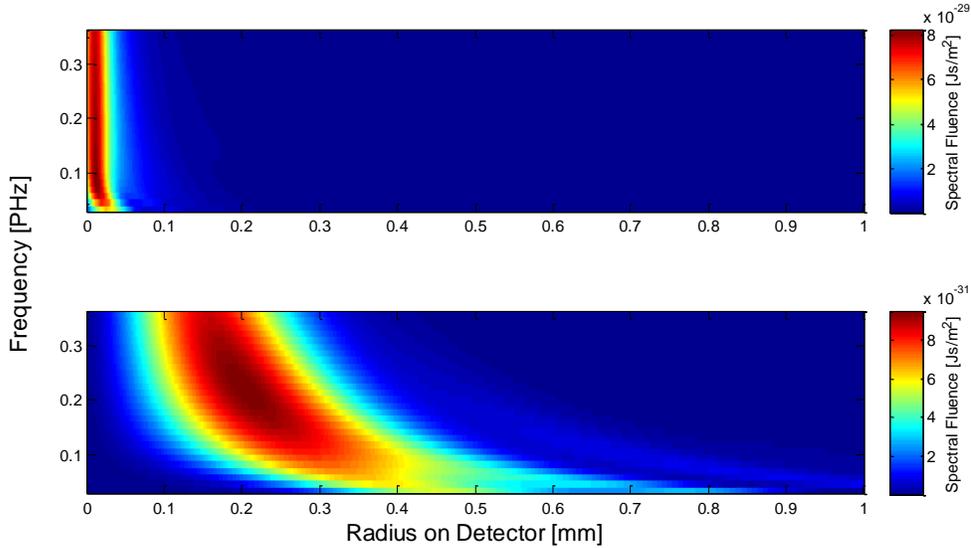

*Figure 4: TR spectral/radial distributions on the 2x2 mm square pixel of the TR detector. Top: beam from tape drive. Bottom: beam from pellicle. Note that the color scale of the bottom image is 2 orders of magnitude lower.*

4. Finally, for each shot, the measured electron energy spectrum is binned within 50MeV steps and the denominator of equation (4) is evaluated. Figure 5a shows the computed spectral energy integrated over one pixel of the detector, dependent on electron energy and assuming a distance between the exit of the gas cell and the tape of 50mm. Note that due to the finite transverse size (i.e. radial formfactor) of the electron beam on the two radiator foils, the emission of high frequency radiation into a large angle is suppressed by destructive interference, i.e. $F_\perp(\omega, \theta) < 1$, which leads to a reduced total emission of short wavelengths. Figure 5b shows the contribution of CTR produced at the tape and the pellicle and their coherent sum for an electron energy of 525MeV. The CTR signal from the pellicle is weaker compared to that from the tape, because of the increased transverse electron beam size at the position of the pellicle and due to its lack of collimation after the first lens, such that only a part of the radiation is collected by the free aperture of the 2[nd] lens and focused onto the detector.

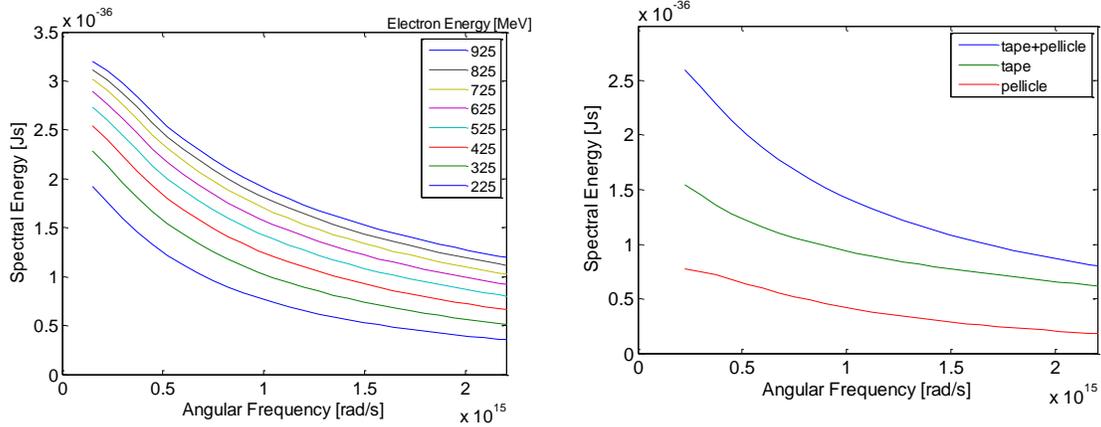

*Figure 5: Computed transmission functions: a) Transmitted signal from both radiators for different mono-energetic electron sheets. b) Transmitted signal from the tape (green) and pellicle (red) itself, and their coherent sum (blue) for an electron energy of 525MeV.*

**Interference effect**

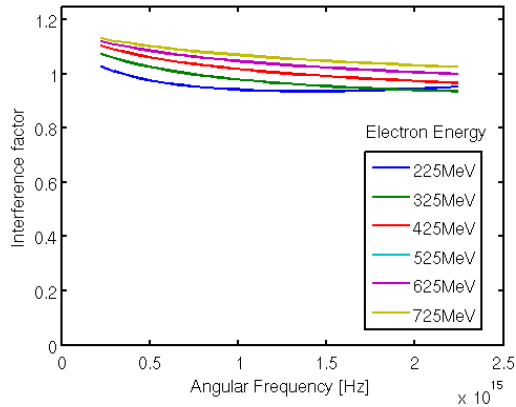

*Figure 6: Interference effect between both radiators in the plane of the detector.*

In order to analyse the interference effect, we have computed the following expression:

$$\text{Interference factor} = \frac{\int (E_1 + E_2)^2 dA}{\int (E_1^2 + E_2^2) dA}$$

Here, $E_1$ and $E_2$ represent the electric field of beams 1 and 2 (from foil and pellicle, respectively). The result is plotted in Fig. 6, where for electron energies in the range 225-725 MeV we see a max. deviation due to interference of ±13% in the relevant frequency range. The effect is more pronounced at higher electron energies due to a better spatial overlap of both beams in the plane of the detector. Note that due to the small absolute phase delay and small spatial overlap of the two beams no oscillatory behaviour with wavelength occurs.
(the effective path length difference is given by $d/(2\gamma^2)$ and thus for $\gamma=1000$ and d=10 cm amounts to 50nm).

Our analysis shows, that despite the contribution from two CTR sources, the spectral energy in the detection plane is a smooth function of frequency, such that the absolute value of $F(\omega)$ can be obtained from the measured CTR spectrum by equation (4), including the effect of interference. The obtained formfactor is then fed to the BUBBLEWRAP[6] algorithm to obtain the bunch profile $\rho_{||}(z)$.